\documentclass[aps,prl,twocolumn,superscriptaddress,preprintnumbers,amssymb]{revtex4}
\usepackage{graphicx}
\usepackage{latexsym}
\usepackage{amssymb}
\usepackage{amsmath}
\usepackage{amsfonts}
\usepackage{bm}
\usepackage{multirow}
\usepackage{color}
\usepackage{hyperref}
\hypersetup{
colorlinks = true,
linkcolor = [rgb]{0.70,0.13,0.13},
citecolor = [rgb]{0.13,0.55,0.13},
urlcolor = [rgb]{0.25, 0.41, 0.88}}

\newcommand{\ii}{\mathrm{i}}
\newcommand{\dsZ}{\mathbb{Z}}
\newcommand{\Tr}{\operatorname{Tr}}

\renewcommand{\Re}{\operatorname{Re}}

\newcommand{\norm}[1]{{\lVert #1\rVert}}

\newcommand{\eq}[1]{\begin{equation}#1\end{equation}}
\newcommand{\eqs}[1]{\begin{equation}\begin{split}#1\end{split}\end{equation}}
\newcommand{\refcite}[1]{Ref.\,\onlinecite{#1}}
\newcommand{\eqnref}[1]{Eq.~\ref{#1}}
\newcommand{\figref}[1]{Fig.~\ref{#1}}
\newcommand{\tabref}[1]{Tab.~\ref{#1}}

\newcommand{\J}{\tilde J}

\begin{document}

\title{Out-of-Time-Order Correlation in Marginal Many-Body Localized Systems}
\author{Kevin Slagle}
\author{Zhen Bi}
\affiliation{Department of Physics, University of California,
Santa Barbara, CA 93106, USA}
\author{Yi-Zhuang You}
\affiliation{Department of Physics, University of California, Santa Barbara, CA 93106, USA}
\affiliation{Department of Physics, Harvard University, Cambridge, MA 02138, USA}
\author{Cenke Xu}
\affiliation{Department of Physics, University of California,
Santa Barbara, CA 93106, USA}

\date{\today}

\begin{abstract}

We show that the out-of-time-order correlation (OTOC) $ \langle
W(t)^\dagger V(0)^\dagger W(t)V(0)\rangle$ in many-body localized
(MBL) and marginal MBL systems can be efficiently calculated by
the spectrum bifurcation renormalization group (SBRG). We find
that in marginal MBL systems, the scrambling time $t_\text{scr}$
follows a stretched exponential scaling with the distance $d_{WV}$
between the operators $W$ and $V$: $t_\text{scr}\sim
\exp(\sqrt{d_{WV}/l_0})$, which demonstrates Sinai diffusion of
quantum information and the enhanced scrambling by the quantum
criticality in non-chaotic systems.

\end{abstract}

\maketitle


The out-of-time-order correlation
(OTOC)\cite{Larkin:1969kn,Shenker:2014fk,Kitaev:2014it,Shenker:2014ij,Shenker:2015ph,Roberts:2015oq,Roberts:2015fc}
was recently proposed to quantify the scrambling and the butterfly
effect in quantum many-body dynamics, and has attracted great
research interests in quantum
gravity\cite{Kitaev:2015uj,Maldacena:2016hh,Maldacena:2016bx,Maldacena:2016xy},
quantum information\cite{Hosur:2016fy} and condensed
matter\cite{Gu:2016hi,Fu:2016hs,Swingle:2016oe,Zhu:2016cr,Yao:2016xp}
communities. Consider two local unitary operators $W$ and $V$,
along with the many-body Hamiltonian $H$ of the system; the OTOC
is defined as \eq{F(t) = \langle W(t)^\dagger V(0)^\dagger
W(t)V(0)\rangle,} where $W(t)=e^{\ii H t}We^{-\ii H t}$ and
$V(0)=V$ are the operators at time $t$ and time $0$ respectively.
The notation $\langle\cdots\rangle$ stands for either the
expectation value on a pure state of interest (typically a
short-range entangled state), or the ensemble average over a mixed
state density matrix.
The OTOC is closely related to the squared
commutator\cite{Shenker:2014fk,Kitaev:2014it,Maldacena:2016xy} of
the operators: $C(t)=\langle\left|[W(t),V]\right|^2\rangle=2(1-\Re
F(t))$. If $W$ are $V$ are far apart local operators, then their
squared commutator $C(t)$ should vanish initially (for $t=0$). As
time evolves, the operator $W(t)$ will grow in size and complexity,
and eventually spread to the location of the operator $V$, at
which point $C(t)$ develops a finite value. So the growth of the
squared commutator $C(t)$, or the decay of the OTOC $F(t)$,
characterizes the growth of local operators and the spreading of
quantum information, which is a phenomenon known as
scrambling\cite{Page:1993fv,Hayden:2007uf,Sekino:2008pv,Lashkari:2013mj}.
Typically, the OTOC will remain large until the scrambling
time\cite{Sekino:2008pv} $t_\text{scr}$ and decay rapidly once $t
> t_\text{scr}$. The scrambling time $t_\text{scr}$ generally
depends on the distance $d_{WV}$ between $W$ and $V$ operators. If
we treat $W$ as a perturbation to the system, then the OTOC also
describes how a local perturbation $W$ spreads to affect the
measurement of $V$ at a distance $d_{WV}$, which can be viewed as
a quantum version of the butterfly effect.\cite{Shenker:2014fk}
The function $t_\text{scr}(d_{WV})$ describes the onset of the
butterfly effect in space-time and traces out the boundary of the
butterfly light-cone.

Although the OTOC was originally proposed to diagnose quantum
chaos, recently there has been a growing interest to study the
OTOC in non-chaotic quantum many-body systems as well, such as in
rational conformal field theories\cite{Caputa:2016ct,Gu:2016hi}
and in many-body localized (MBL)
systems\cite{Huang:2016wk,Fan:2016oy,Swingle:2016rw,Chen:2016ec,He:2016bf,Chen:2016bx}.
In MBL
systems,\cite{Avishai:1996qe,Avishai:1997cs,GMP:2005,BAA:2006,Imbrie:2014,Huse:2015rv}
energy, charge, and other local conserved quantities can not
defuse due to the localization of excitations in the presence of
strong disorder. Nevertheless, the quantum information can still
propagate, as first demonstrated by the unbounded growth of
entanglement\cite{Moore:2012ge,DeChiara:2006zf,Znidaric:2008he,Abanin:2013ta,Nanduri:2014jh}
after a global quench. The propagation of quantum information in
MBL systems was also observed from OTOC
measurements.\cite{Huang:2016wk,Fan:2016oy,He:2016bf} Compared to
quantum chaotic systems, where the scrambling time scales linearly
with the spatial separation between the operators
$t_\text{scr}\sim d_{WV}$, MBL systems were found to be a much
slower scrambler with a scrambling time that scales exponentially
with the operator separation $t_\text{scr}\sim
\exp({d_{WV}/\xi})$. On the other hand, it was
conjectured\cite{Shen:2016ov} that quantum critical fluctuations
can enhance scrambling in chaotic systems. In this work, we found
that criticality also enhances scrambling in non-chaotic MBL
systems. We will demonstrate that the scrambling time follows a
stretched exponential scaling $t_\text{scr}\sim
\exp(\sqrt{d_{WV}/l_0})$ for marginal MBL
systems\cite{Potter:2014mh} (i.e.\,quantum critical MBL systems),
which is different from both the quantum chaotic and the MBL
behaviors mentioned above.

We used the spectrum bifurcation renormalization group
(SBRG)\cite{You:2015sb,Slagle:2016mk} approach to calculate the
OTOC in MBL and marginal MBL systems. SBRG is an efficient
numerical approach to construct the MBL effective Hamiltonian from
a given disordered quantum many-body Hamiltonian. The idea of SBRG
is similar to the real space renormalization group for excited
states
(RSRG-X)\cite{Altman:2013rg,Swingle:2013oy,Refael:2013du,Altman:2014hg,Potter:2015th,Vasseur:2015ys}.
At each RG step, the leading energy scale term $H_0$ in the
Hamiltonian is identified and the whole Hamiltonian is rotated to
the (block) diagonal basis of $H_0$; then the terms in the
off-diagonal blocks are reduced by the 2nd order perturbation.
SBRG uses Clifford gates to boost the calculation efficiency for
qubit models, such that the full spectrum is obtained in one run
of RG (in contrast to RSRG-X which targets a single eigenstate at
a time). With SBRG we can push the calculation of OTOC to much
larger system size (e.g.\;256 spins in this work) than exact
diagonalization, hence verifying the scaling behaviors of the
butterfly light-cones over a much larger scale.


We start by deriving the formula for the OTOC that can be used in
the SBRG calculations. The output of SBRG\cite{You:2015sb} is the
MBL effective
Hamiltonian,\cite{Abanin:2013ta,Swingle:2013oy,Abanin:2013lc,Huse:2014ec,Kim:2014zj,Abanin:2015io,Rademaker:2015ve}
which can be written in terms of the stabilizers (l-bits)
$\tau^z_a$ ($a=1,2,\cdots,L$ labels the stabilizers) as,
\eq{H_\text{MBL}=\sum_{a}\epsilon_{a}
\tau^z_a+\sum_{a,b}\epsilon_{ab}\tau^z_a\tau^z_b+\sum_{a,b,c}\epsilon_{abc}\tau^z_a\tau^z_b\tau^z_c\cdots,
\label{eq: H_MBL}} which contains single-body terms
$\epsilon_a\tau^z_a$, two-body terms
$\epsilon_{ab}\tau^z_a\tau^z_b$ and higher-body terms. The key
difference between Anderson and MBL insulators is that the
two-body and higher-body terms are absent in the former while
present in the later. $H_\text{MBL}$ can also describe the
marginal MBL system, where the major modification is that the
stabilizers $\tau^z_a$ will be quasi-long-ranged (The chance of
finding a stabilizer decays as a power-law with its length),
instead of exponentially localized in the MBL system.

The stabilizers all commute with each other and also commute with
the Hamiltonian, i.e. $[\tau^z_a,\tau^z_b]=[\tau^z_a,H_\text{MBL}]=0$.
To simplify the notation, we denote each product of stabilizers as
$\tau^z_{abc\cdots}=\tau^z_a\tau^z_b\tau^z_c\cdots$. We may further bundle
the subscript indices together and write
\eq{H_\text{MBL}=\sum_{A}\epsilon_{A}\tau^z_{A},} where
$A=abc\cdots$ stands for a sequence of stabilizer indices and
$\tau^z_A=\prod_{a\in A}\tau^z_a$. Since the Hamiltonian
$H_\text{MBL}$ is a sum of commuting terms $\epsilon_{A}\tau^z_{A}$,
the time-evolution operator $U(t)$ can be factorized to the
product of unitary operators generated by every $\tau^z_{A}$
operator independently, \eq{U(t)=e^{-\ii t
H_\text{MBL}}=\prod_{\tau^z_A}e^{-\ii t \epsilon_{A} \tau^z_{A}}.} The
product runs over all $\tau^z_A$ operators in the Hamiltonian
$H_\text{MBL}$. The time-dependent operator $W(t)$ is then given
by $W(t)=U(t)^\dagger WU(t)$.

We begin by transforming the $W$ and $V$ operators to the $\tau^z$
basis, which involves a unitary transformation composed of
Clifford rotations with perturbative Schrieffer-Wolff corrections
\cite{You:2015sb}. In this basis, $W$ and $V$ are then linear
combinations of products of Pauli operators. For simplicity, we
will neglect the perturbative Schrieffer-Wolff corrections, which
are small in the limit of large disorder. In
Ref.~\onlinecite{You:2015sb} we tested this approximation by
restoring the many-body wave function from the Clifford rotation
only, and benchmarking the result with exact diagonalization. Good
wave function fidelity is achieved as long as the disorder is
strong. We will only consider $W$ and $V$ which are products of
Pauli operators (called \emph{Pauli strings}) in the physical
basis, which implies that they will remain Pauli strings in the
$\tau^z$ basis (since Clifford rotations map Pauli strings to
Pauli strings). Therefore their algebraic relations with
$\tau^z_A$ is rather simple: $W$ and $V$ can either commute or
anticommute with $\tau^z_A$~\footnote{In MBL phase or marginal MBL
phase, $W$ and $V$ should be an infinite sum of Pauli strings in
the $\tau$ basis, with coefficients decaying exponentially with
both the disorder and the range~\cite{Huse:2014ec,Huse:2015cd},
see for instance Fig.3 of Ref.~\onlinecite{Slagle:2016mk}.
Neglecting the Schrieffer-Wolff correction keeps the leading Pauli
string in this expansion.}. Let $\mathcal{C}_W$ (or
$\mathcal{A}_W$) be the set of $\tau^z_A$ that \emph{commute} (or
\emph{anticommute}) with $W$. Any $\tau^z_A$ in $H_\text{MBL}$
either belongs to $\mathcal{C}_W$ or $\mathcal{A}_W$ for any given
$W$. With this setup, we can calculate the time-evolution of $W$
as follows \eqs{\label{eq: W(t)}
W(t)&=\prod_{\tau^z_A}e^{\ii t \epsilon_{A} \tau^z_{A}}W\prod_{\tau^z_A}e^{-\ii t \epsilon_{A} \tau^z_{A}}\\
&=W\prod_{\tau^z_A\in\mathcal{A}_W}e^{-2\ii t \epsilon_{A}
\tau^z_{A}}.} The unitary operators generated by
$\tau^z_A\in\mathcal{C}_W$ will annihilate each other by commuting
through $W$, so only those generated by $\tau^z_A\in\mathcal{A}_W$
will survive. Suppose $W$ is a local operator (e.g. an on-site
Pauli operator); then \eqnref{eq: W(t)} indicates that the support
(or the size) of $W(t)$ will grow in time. $W(t)$ starts out with
$W(0)=W$ initially, and as time evolves, $W$ will expand via a
product of non-local operators  $e^{-2\ii t \epsilon_{A}
\tau^z_{A}}$. Each of them gradually evolves from $1$ to $\ii
\tau^z_{A}$ in the time scale $\sim\epsilon_{A}^{-1}$. $\tau^z_{A}$
terms that are more non-local typically have smaller energy scales
$\epsilon_{A}$ in the local Hamiltonian $H_\text{MBL}$, and thus
take longer time to contribute to $W(t)$. So the operator $W(t)$
will grow gradually. Accordingly, as $W(t)$ becomes non-local, the
quantum information associated with  $W$ will be spread throughout
the system and can not be retrieved by local measurements, which
illustrates the idea of quantum
chaos\cite{Srednicki:1994ns,Rigol:2010ls,Roberts:2015fc,Hosur:2016fy}
and scrambling\cite{Page:1993fv,Hayden:2007uf,Sekino:2008pv}.

The OTOC was proposed to quantify the growth of the operator and
the scrambling effect. Here let us discuss the OTOC at ``infinite
temperature'' where the density matrix of the system is simply
identity, so that \eq{F(t) = \Tr W(t)V(0)W(t)V(0),\label{eq:OTOC}}
(the daggers are omitted as we assume both $W$ and $V$ are
Hermitian Pauli operators). Following the similar calculation in
\eqnref{eq: W(t)}, we find \eq{F(t)=\Tr
WVWV\prod_{\tau^z_A\in\mathcal{A}_W\cap\mathcal{A}_V}e^{4\ii
t\epsilon_A\tau^z_A},} where $\tau^z_A$ are the terms in
$H_\text{MBL}$ that anticommute with both $W$ and $V$. As both $W$
and $V$ are Pauli strings, regardless of whether they commute or
anticommute, this just amounts to an overall sign in $WVWV=\pm 1$,
which is not important. So we might as well assume $[W,V]=0$
(which is the case for far apart local operators), then the OTOC
simply reads
$F(t)=\Tr\prod_{\tau^z_A\in\mathcal{A}_W\cap\mathcal{A}_V}e^{4\ii
t\epsilon_A\tau^z_A}$. The unitary operators can be expanded, i.e.
\eq{F(t)=\Tr\prod_{\tau^z_A\in\mathcal{A}_W\cap\mathcal{A}_V}(\cos(4\epsilon_A
t)+\ii\tau^z_A\sin(4 \epsilon_A t)).} We take an approximation by
dropping all the $\sin(4 \epsilon_A t)$ terms in the expansion (to
be justified shortly),\cite{Fan:2016oy,Chen:2016ec} and arrive at
a simple formula for the OTOC of MBL and marginal MBL systems,
\begin{equation}\label{eq: F(t) cos} F(t)\simeq
\prod_{\tau^z_A\in\mathcal{A}_W\cap\mathcal{A}_V} \cos(4
\epsilon_A t).\end{equation} In numerics, we first run the SBRG on
a given quantum many-body Hamiltonian to generate the MBL
effective Hamiltonian $H_\text{MBL}$. From $H_\text{MBL}$ we
filter out all terms $\epsilon_A\tau^z_A$ that anticommute with
both $W$ and $V$ (recall that $W$ and $V$ are Pauli strings in the
$\tau^z_A$ basis since we dropped the perturbative
Schrieffer-Wolff corrections) and collect their energy
coefficients $\epsilon_A$. Then the OTOC can be evaluated very
efficiently according to \eqnref{eq: F(t) cos}. We must bear in
mind that \eqnref{eq: F(t) cos} does not apply to the thermalized
system, because our starting point, the MBL effective Hamiltonian
$H_\text{MBL}$, breaks down in the thermalized phase.

The approximation we made in \eqnref{eq: F(t) cos} is to drop all
terms in the expansion that contain the product of
$\sin(4\epsilon_A t)$. Such terms will only arise when several
different $\tau^z_A$ operators product to identity so as to survive
the trace. However note that all $\tau^z_A$ in \eqnref{eq: F(t) cos}
are taken from the set $\mathcal{A}_W\cap\mathcal{A}_V$, within
which one must have at least four $\tau^z_A$ product together to
reach the identity (as long as $W$ and $V$ commute). Thus the
$\sin(4\epsilon_A t)$ factors appear as products of four or more,
whose short-time behavior is suppressed by $\sim t^4$ (as
$t\to0$). In conclusion, such terms will never dominate the
expansion until after $t_\text{scr}$.

To calculate the OTOC of more general operators $W$ and $V$, or to
include the perturbative Schrieffer-Wolff corrections, one can
expand the operators as a sum of Pauli strings in the $\tau^z$
basis, and express the OTOC as a sum of these operators. In the
following, we will only focus on the OTOC of Pauli strings without
Schrieffer-Wolff corrections.



The OTOC starts out at 1 and decays to 0. The time for the onset
of the decay is defined as the scrambling time
$t_\text{scr}$.\cite{Sekino:2008pv,Swingle:2016oe} One can
estimate the scrambling time based on \eqnref{eq: F(t) cos}. At
short-time, $\cos(4\epsilon_A t)$ can be Taylor expanded to
$1-\frac{1}{2}(4\epsilon_A t)^2+\cdots$, so the OTOC behaves as
\eq{F(t)=1-\frac{1}{2}(4\norm{\epsilon_A}_{W,V} t)^2+\cdots,}
where the energy scale $\norm{\epsilon_A}_{W,V}$ is defined via
\eq{\norm{\epsilon_A}_{W,V}^2=\sum_{\tau^z_A\in\mathcal{A}_W\cap\mathcal{A}_V}\epsilon_A^2.}
So the scrambling time $t_\text{scr}$ is set by this energy scale
as $t_\text{scr}=\norm{\epsilon_A}_{W,V}^{-1}$. The energy scale
$\norm{\epsilon_A}_{W,V}$ is not just a Frobenius norm of the
energy coefficients in $H_\text{MBL}$, it also sensitively depends
on the operators $W$ and $V$. If $W$ and $V$ are local operators,
then the scrambling time scales only with the distance $d_{WV}$
between $W$ and $V$. The scaling behavior can be used to
distinguish Anderson localization, MBL, marginal MBL and
ergodic\cite{Hartnoll:2015ef,Blake:2016ee,Blake:2016xw,Gu:2016zm}
systems, as concluded in \tabref{tab: scaling}.

\begin{table}[htbp]
\caption{Scaling of scrambling time $t_\text{scr}$ with the
operator distance $d_{WV}$ as $d_{WV}\to\infty$ in different types
of systems.}
\begin{center}
\begin{tabular}{c|cccc}
& Anderson & MBL & Marginal MBL & Ergodic\\
\hline
$\ln t_\text{scr}$ & $\infty$ & $\sim d_{WV}$ & $\sim d_{WV}^{1/2}$ & $\ln d_{WV}$
\end{tabular}
\end{center}
\label{tab: scaling}
\end{table}

For Anderson insulators, if the spacial separation between $W$ and
$V$ is much greater than the localization length, then
$\mathcal{A}_W\cap\mathcal{A}_V$ is usually an empty set, i.e.
there is almost no stabilizer that can anticommute with both $W$
and $V$ because all stabilizers are exponentially localized within
the localization length. (Recall that all terms in $H_\text{MBL}$
(\eqnref{eq: H_MBL}) are stabilizers for Anderson insulators, and
that we're approximating the stabilizers as a product of Pauli
operators by neglecting the perturbative corrections.) In this
case $\norm{\epsilon_A}_{W,V}\to 0$ and hence $t_\text{scr}\to
\infty$. So the OTOC will remain finite and not decay in time for
far apart $W$ and $V$, meaning that there is no scrambling in
Anderson insulators.

The situation is different if we add interactions. For MBL
systems, far apart $W$ and $V$ operators can be connected by
many-body interaction terms in $H_\text{MBL}$. A typical
contribution comes from the two-body terms
$\epsilon_{ab}\tau^z_{a}\tau^z_{b}$ with $\tau^z_a$ localized
around $W$ and $\tau^z_b$ localized around $V$. Then
$\norm{\epsilon_A}_{W,V}\simeq\norm{\epsilon_{ab}}\sim
e^{-x_{ab}/\xi}$, where $x_{ab}$ is the distance between
$\tau^z_a$ and $\tau^z_b$, which is also roughly the distance
$d_{WV}$ between $W$ and $V$. So the scrambling time
$t_\text{scr}$ follows $\ln t_\text{scr}\sim d_{WV}/\xi$, leading
to a logarithmic butterfly light-cone in the MBL
system.\cite{Huang:2016wk,Fan:2016oy,Swingle:2016rw,Chen:2016ec,He:2016bf}

Another direction out of Anderson insulators is to consider
quantum critical systems, i.e. marginal MBL systems. In these
systems, each stabilizer $\tau^z_a$ itself becomes power-law quasi
localized, and can connect spatially far separated $W$ and $V$.
Then the energy scale can be dominated by the single-body energy
$\norm{\epsilon_A}_{W,V}\simeq\norm{\epsilon_{a}}\sim
e^{-\sqrt{l/l_0}}$, which follows the ``stretched exponential''
scaling with respect to the length $l$ of the stabilizer (where
$l_0$ is a length scale depending on the initial disorder
strength). This scaling is an exact result in the free limit by
RSRG and has been shown to apply to interacting cases in
\figref{fig:EvsL} as well in Ref.\,\onlinecite{You:2015sb}. $l$ is
also roughly the distance $d_{WV}$ between $W$ and $V$. Therefore
the scrambling time $t_\text{scr}$ follows $\ln t_\text{scr}\sim
\sqrt{d_{WV}}$, leading to a squared logarithmic butterfly
light-cone in the marginal MBL system. Because the scrambling in
the marginal MBL system is determined by the single-body energy
scale, the butterfly light-cone is not much affected by the
absence or presence of the interaction.

\begin{figure}[htbp]
\includegraphics[width=0.7\columnwidth]{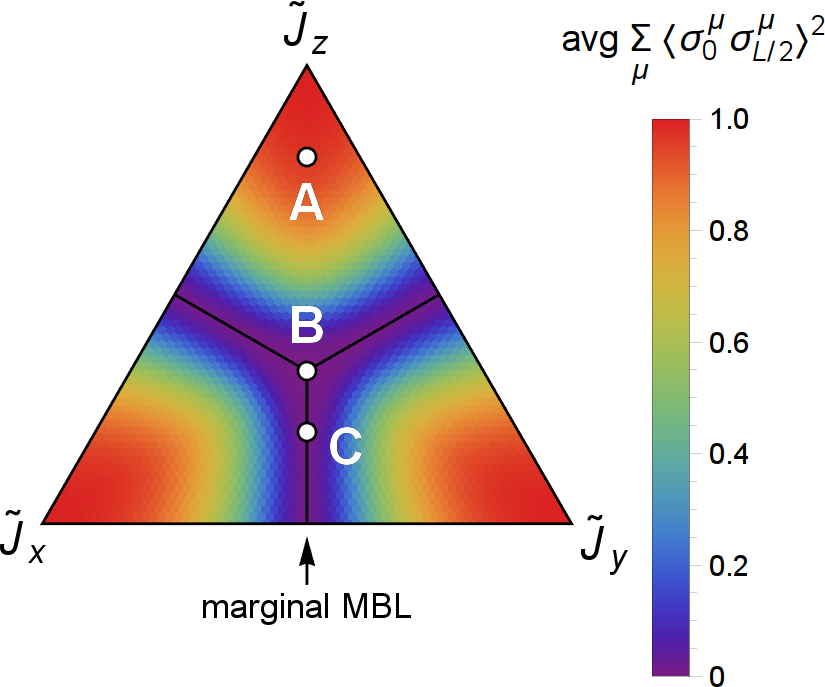}
\caption{ A ternary plot (copied from \cite{Slagle:2016mk}) of the
disorder and energy averaged Edwards-Anderson correlator vs
coupling constants ($0 < \J_{x,y,z} < 1$) for the XYZ spin chain
of length $L=256$. We use this plot to sketch the phase diagram.
When $\J_z > \max(\J_x, \J_y)$, the system is in an MBL $\dsZ_2$
spin glass state. When $\J_z < \J_x = \J_y$, the system is in a
marginal MBL phase. (The other phases are given by permutations of
$x,y,z$.) The white dots correspond to the points in the phase
diagram that are shown in \figref{fig:sqrt OTOC}.
}\label{fig:phase diagram}
\end{figure}

\begin{figure}[htbp]
\includegraphics[width=\columnwidth]{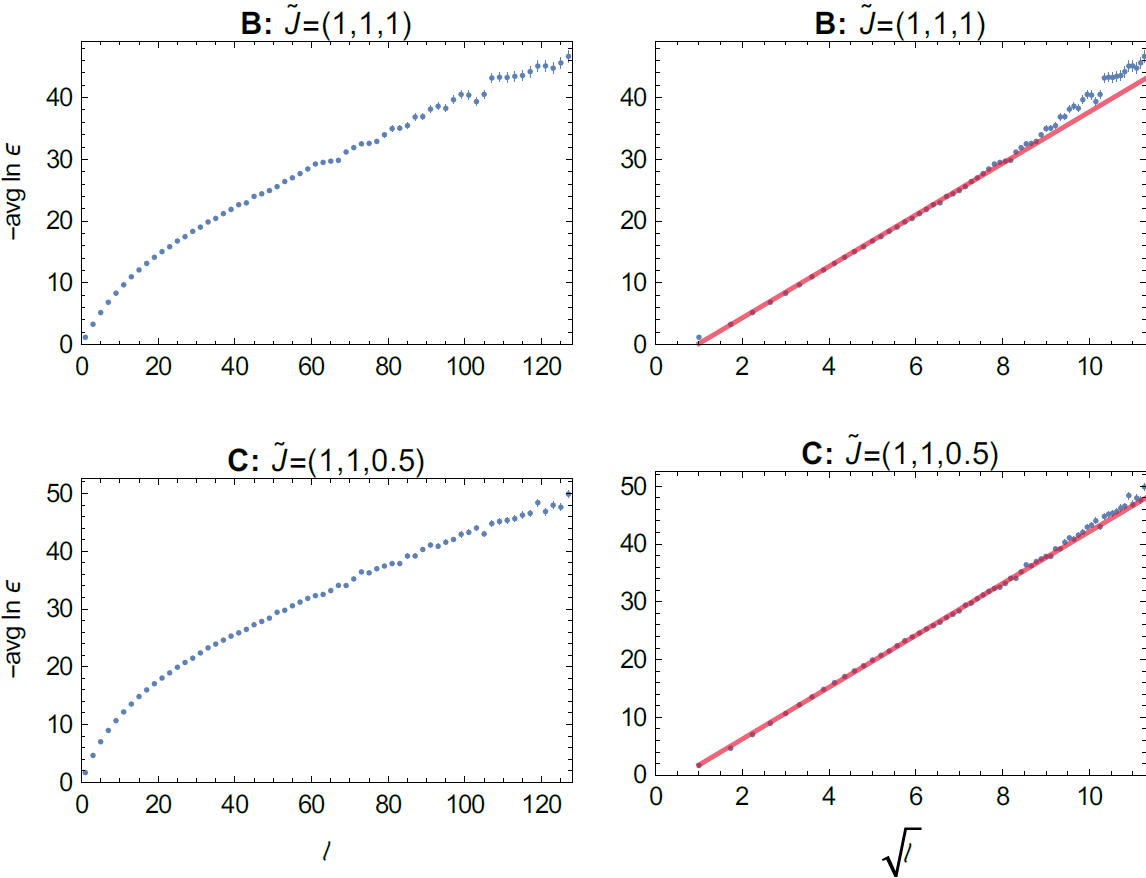}
\caption{ Disorder average of the log of the stabilizer energy
$\ln\epsilon$ vs stabilizer length $\ell$ on a periodic lattice of
256 spins in the marginal MBL phase of the XYZ spin chain. This
figure verifies the scaling $-\ln\epsilon \sim \sqrt{\ell}$. The
two rows (of plots) correspond to different points (B and C) in
the phase diagram \figref{fig:phase diagram} while the two columns
correspond to different horizontal axes: $|i-j$ vs $\sqrt{|i-j|}$.
(The fully MBL point (A) is not shown since it is deep in the MBL
phase where nearly all stabilizers are of very short length.) The
stabilizer length is calculated by writing a stabilizer $\tau_a^z$
in the physical basis and dropping all perturbative
Schrieffer-Wolff corrections. The result is a product of Pauli
operators at different sites. The stabilizer length is then the
length of the shortest continuous sequence of sites (on the
periodic lattice) that contains all of the Pauli operators. (Due
to a slight even-odd effect, only odd stabilizer lengths are
shown. $2^{16}$ disorder samples are used. Error bars denote one
standard deviation statistical errors.) }\label{fig:EvsL}
\end{figure}

\begin{figure}[h]
\includegraphics[width=\columnwidth]{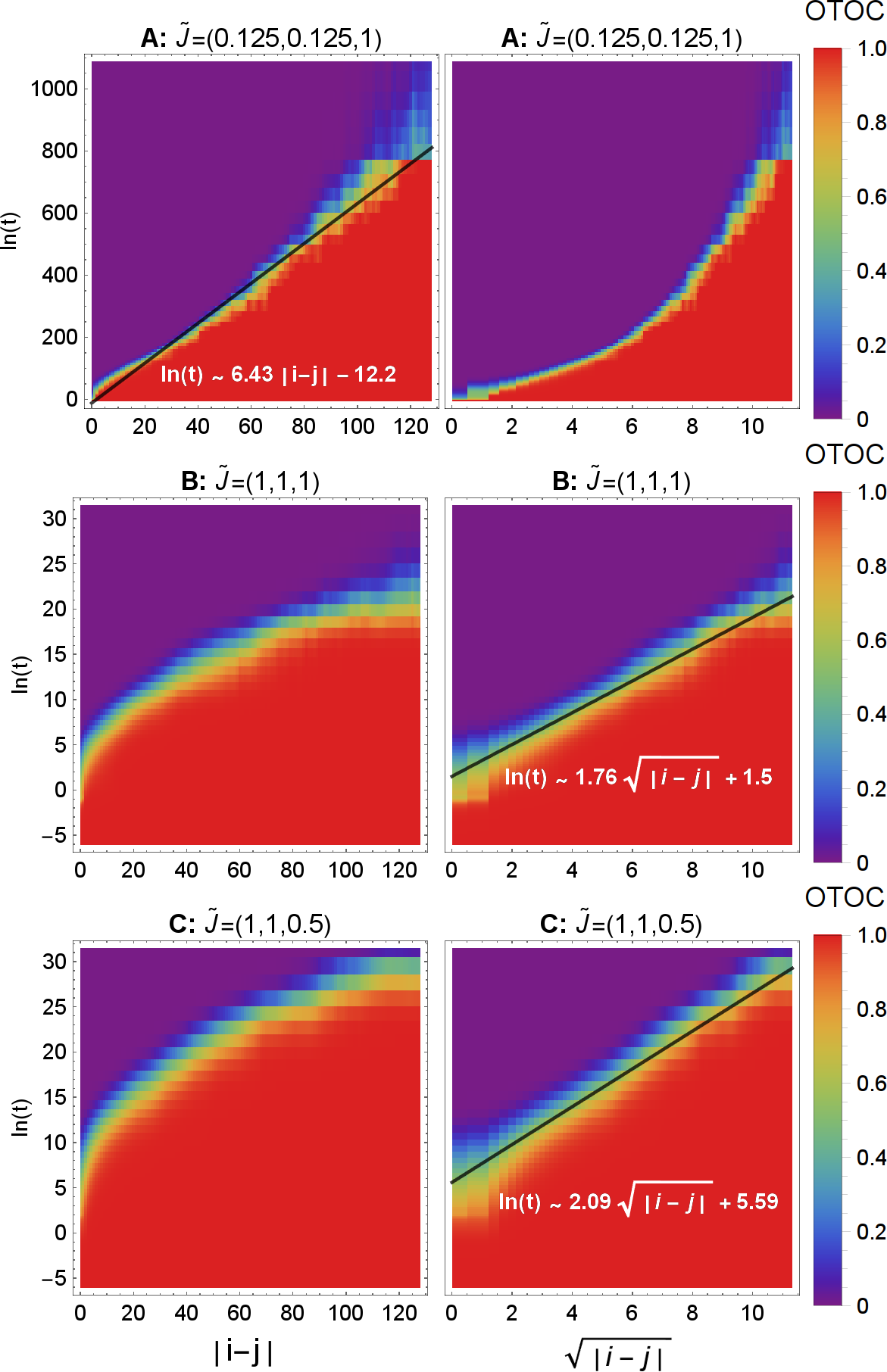}
\caption{ Disorder average (using a geometric mean
(\eqnref{eq:geometric mean})) of the out-of-time-order correlation
(OTOC) (\eqnref{eq:OTOC}) $\exp \text{avg} \ln |F(t)|$ of
$W=\sigma_i^x$ and $V=\sigma_j^y$ showing how the light cone of
the (geometric mean) OTOC depends on the time $t$ and distance
$d_{WV}=|i-j|$ separation of $W$ and $V$ (on a lattice with 256
spins). Specifically, the light cone grows like $t_\text{scr} \sim
\exp(\sqrt{d_{WV}/l_0})$. As in \figref{fig:EvsL}, the rows (of
plots) correspond to different points in the phase diagram
\figref{fig:phase diagram} while the two columns correspond to
different horizontal axes: $|i-j$ vs $\sqrt{|i-j|}$. Fits are
shown for the cases when the scrambling time scaling agrees with
the choice of horizontal axes. As can be seen from
\figref{fig:error OTOC}, the statistical errors and finite system
size do not significantly affect the linear fit. ($2^9$ disorder
samples are used.)}\label{fig:sqrt OTOC}
\end{figure}

\begin{figure}
\includegraphics[width=0.9 \columnwidth]{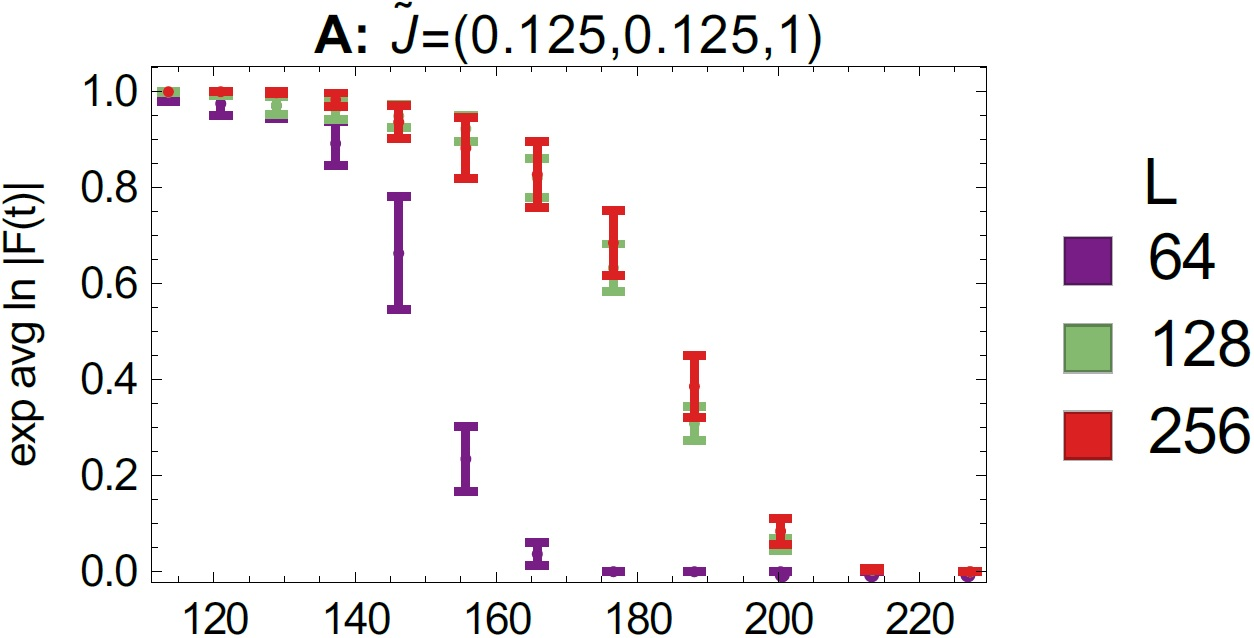}
\includegraphics[width=0.9 \columnwidth]{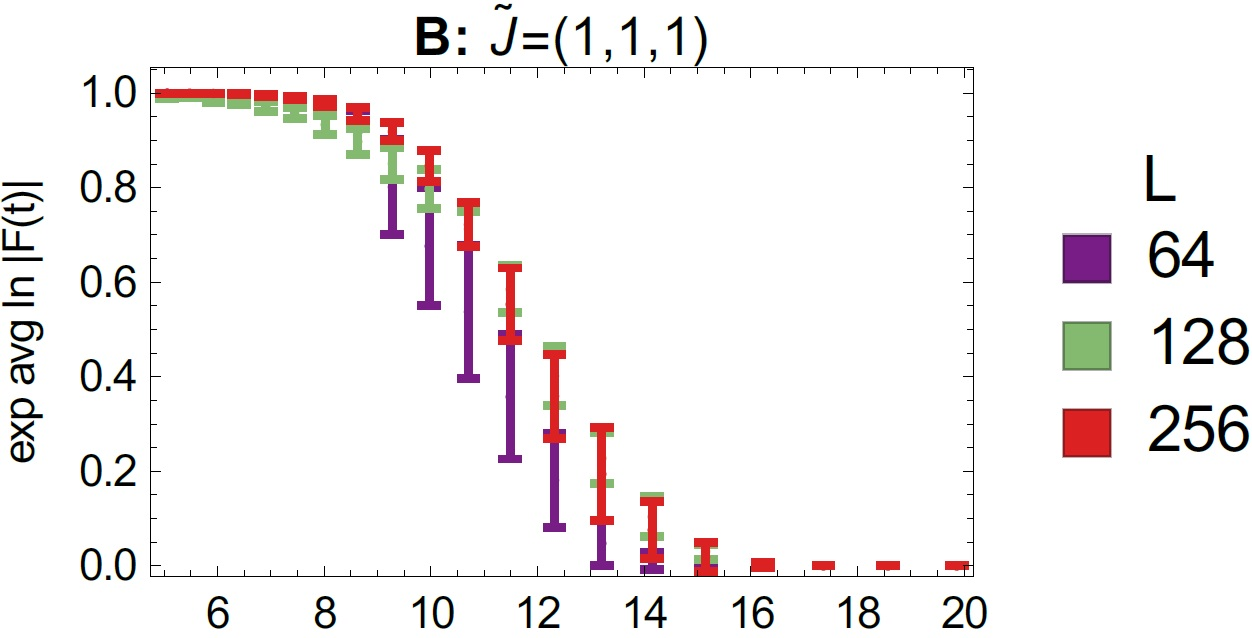}
\includegraphics[width=0.9 \columnwidth]{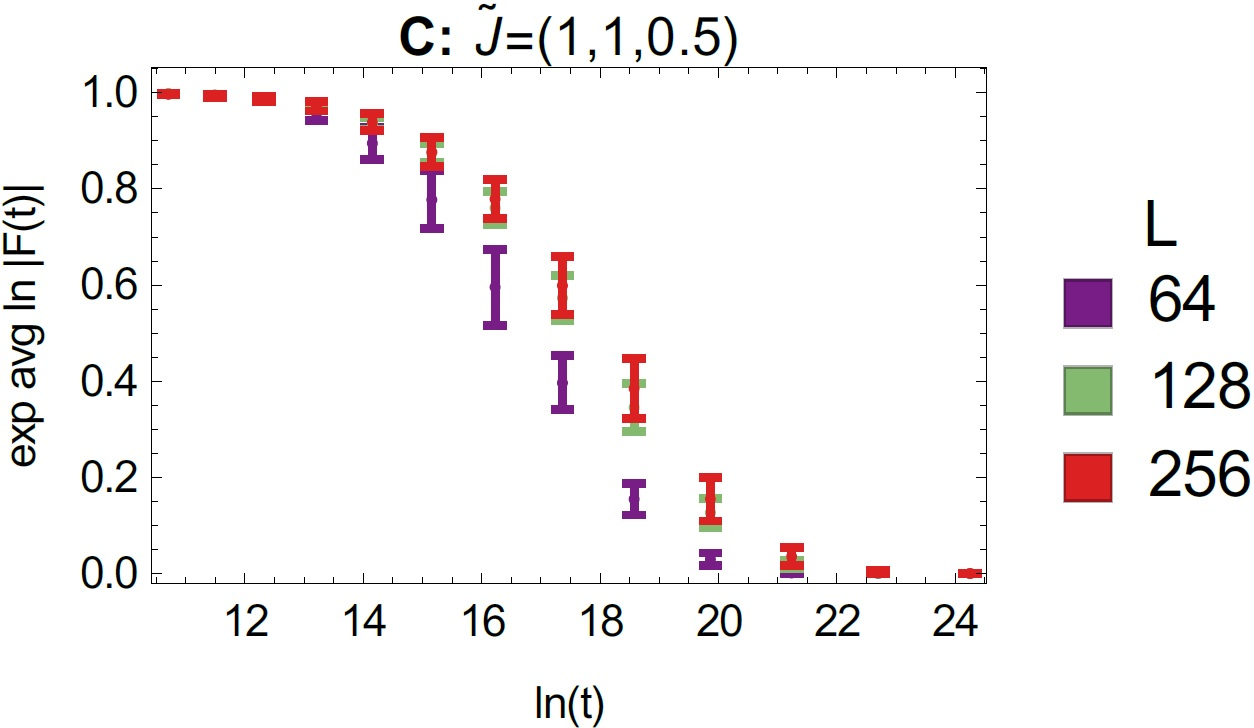}
\caption{ A $d_{WV} = |i-j| = 32$ slice of \figref{fig:sqrt OTOC}
for different system sizes (and the same three points of the phase
diagram \figref{fig:phase diagram}). That is, we plot the disorder
average (using a geometric mean (\eqnref{eq:geometric mean})) of
the out-of-time-order correlation (OTOC) (\eqnref{eq:OTOC}) $\exp
\text{avg} \ln |F(t)|$ of $W=\sigma_i^x$ and $V=\sigma_{i+32}^y$
for system sizes $L=64,128,256$. We see that the OTOC for $|i-j| =
32$ converges very quickly with system size and has essentially
completely converged by $L=128$, which is only four times $|i-j|$.
Thus, we expect \figref{fig:sqrt OTOC} (for which $L=256$) to have
converged for all $|i-j| \leq 256/4 = 64$ (or $\sqrt{|i-j|} \leq
8$). Even when $L=64$, which is only twice $|i-j|$, the OTOC has
already mostly converged. Error bars are statistical errors which
are calculated using the bootstrap method
\cite{ChenErrors,EfronBootstrap} and are small enough to not have
a significant effect on our light cone measurements.
}\label{fig:error OTOC}
\end{figure}

To verify the above theoretical proposals, we numerically measure
the OTOC in MBL and marginal MBL systems by SBRG. The model we
study is the XYZ spin chain with strong disorder on a periodic 1D
lattice. The Hamiltonian is given by\cite{Slagle:2016mk}
\eq{H=\sum_{i=1}^{L}(J_{i,x}\sigma_i^x\sigma_{i+1}^x+J_{i,y}\sigma_i^y\sigma_{i+1}^y+J_{i,z}\sigma_i^z\sigma_{i+1}^z),}
where $\sigma_i^\mu$ ($\mu=x,y,z$) are the spin operators on $i$th
site of a 1D lattice of length $L=256$. The random couplings
$J_{i,\mu}\in[0,J_\mu]$ are independently drawn from the power-law
distribution $\mathrm{PDF}(J_{i,\mu}) = 1/(\Gamma
J_{i,\mu})(J_{i,\mu}/J_{\mu})^{1/\Gamma}$, where $0 < \Gamma <
\infty$ controls the disorder strength. We define
\begin{equation}
\J_\mu \equiv J_\mu^{1/\Gamma},\end{equation} and take $\J= (\J_x,
\J_y, \J_z)$ as the tuning parameters. In this work, a large
disorder strength of $\Gamma=4$ was used\refcite{Slagle:2016mk}.
The model has three spin glass MBL phases corresponding to the
large $\J_x$, $\J_y$ or $\J_z$ limits respectively, as shown in
\figref{fig:phase diagram}, where the spin flip
$\dsZ_2\times\dsZ_2$ symmetry is broken in every many-body
eigenstate of the Hamiltonian.
The spin glass phases are separated by three phase boundaries,
where all the eigenstates become quantum critical, and the system
is at the marginal MBL point.

We will focus along the line of $\J_x=\J_y$ and study the behavior
of OTOC by SBRG. Details of the SBRG algorithm are given in
\refcite{You:2015sb, Slagle:2016mk}. In short, SBRG can accurately
simulate phases with spectrum bifurcation in the limit of large
disorder where the Hamiltonian is written as a sum of products of
Pauli operators where each product of Pauli operators has an
independently random coefficient. By large disorder, we mean that
for every coefficient $h_i$, the standard deviation of $\log(h_i)$
is large. SBRG performs well in both the fully MBL and the
marginal MBL phases. SBRG does not perform well in or near thermal
phases. In this work we keep the largest 1024 additional terms
during each RG step. \footnote{That is, in this work we keep 1024
(instead of only 256 as in \refcite{Slagle:2016mk}) additional
terms to $\Sigma^2$; see Appendix B in \refcite{Slagle:2016mk}.}
Keeping more terms in $\Sigma^2$ was beneficial for this work
because it allows SBRG to capture more of the small terms in
$H_\text{MBL}$ (\eqnref{eq: H_MBL}), which allows us to more
accurately calculate the OTOC at larger distances in
\figref{fig:sqrt OTOC}.

In \figref{fig:sqrt OTOC} we show the color plots of the OTOC
$F(t)=\Tr W(t)V(0)W(t)V(0)$ for local operators $W=\sigma_i^x$ and
$V=\sigma_j^y$ at sites $i$ and $j$ respectively. The choice of
the operators is quite generic. The primary consideration is to
avoid the operators that commute with most of the local integral
of motions (LIOMs) in the MBL system, else it is be difficult to
observe the decay of the OTOC within reasonable time
scale.\cite{Huang:2016wk} As the LIOMs in the large-$\tilde{J}_z$
spin glass phase are mainly $\sigma_i^z\sigma_{i+1}^z$, we will
not choose $W$ or $V$ to be $\sigma^z$ operators. Other than that,
we have tried several different choices of $W$ and $V$, and the
resulting OTOC is similar to what is shown in \figref{fig:sqrt
OTOC}.

In our calculation, the disorder averaging is done using a
geometric mean (which measures the typical value of the OTOC).
More specifically, we calculated
\begin{equation}
\label{eq:geometric mean} \exp \text{avg} \ln |F(t)| = \exp\left(
\frac{1}{N_\delta} \sum_\delta \ln |F(t)| \right)
\end{equation}
where $\sum_\delta$ denotes the summation over $N_\delta$ disorder
samples. The typical correlation function in a marginal MBL phase,
and its crucial difference from the arithmetic mean value (often
dominated by rare events) was discussed in many previous
studies~\cite{Fisher:1992is,Fisher:1994he,Fisher:1995cr,Huse:2001ez}
In our case, the geometric mean was used because it was not
possible to accurately calculate the ordinary mean at large time
and distance separation using our SBRG methods. For a given
disorder sample, sometimes SBRG does not manage to find enough
terms in \eqnref{eq: F(t) cos}, which results in an $F(t)$ that is
too large at large time $t$. This error can substantially affect
the arithmetic mean of $|F(t)|$, but is negligible in the
geometric mean. Therefore we use the typical OTOC (i.e. geometric
mean) to reduce the rare-event effect.

We see in \figref{fig:sqrt OTOC} that in general the OTOC starts
out from 1 and decays to 0. The time-scale for the onset of the
decay, i.e.\,the scrambling time $t_\text{scr}$, grows
monotonically with the distance $d_{WV}=|i-j|$ between the
operators $W$ and $V$. The top row of \figref{fig:sqrt OTOC} is
deep in the MBL spin glass phase with $\J_z/\J_{x,y} = 8$, while
the middle and bottom row are at the marginal MBL critical points
with $\J_x=\J_y=\J_z$ or $2\J_z$ (see \figref{fig:phase diagram}
for a phase diagram). The left column is plotted with $|i-j|$ as
the horizontal axis, while the right column uses $\sqrt{|i-j|}$.
The side-by-side comparison shows that in the MBL phase, the OTOC
light cone is logarithmic $\ln t_\text{scr} \sim d_{WV}$, while
the marginal MBL light cone obeys $\ln t_\text{scr} \sim
d_{WV}^{1/2}$, as expected. On the other hand, if we treat
$d_{WV}$ as a function of time: \eq{\label{eq:
Sinai}d_{WV}\sim(\ln t)^2,} then $d_{WV}$ can be viewed as the
size of the operator $W(t)$. So \eqnref{eq: Sinai} also describes
the slow spreading of the quantum information of operator $W$ in
the system. Its transport universality class is known as the Sinai
diffusion,\cite{Sinai:1982sf} which governs the transport in
critical Anderson localized system.\cite{Bagrets:2016kx} Our
calculation demonstrates that the spreading of quantum information
in marginal MBL systems also follows the Sinai diffusion rule.
Interaction does not seems to affect the diffusion behavior,
probably because the operator growth in the marginal MBL system is
dominated by the single-body terms $\sum_a\epsilon_a\tau^z_a$ of
the MBL Hamiltonian. The Sinai diffusion of quantum information is
also seen in the entanglement growth $S(t)\sim (\ln t)^2$ for
Ising-like marginal MBL systems, as studied in
Ref.\,\onlinecite{Potter:2015th,Altman:2013rg,Altman:2014dq}.

In summary, we demonstrated how the OTOC in MBL and marginal MBL
systems can be efficiently calculated using the SBRG approach. The
system size can be pushed to several hundred sites, much larger
than the previous exact diagonalization studies. We confirmed the
logarithmic butterfly light cone $\ln t_\text{scr}\sim d_{WV}$ in
the MBL system. We found the marginal MBL system is a faster
scrambler due to quantum criticality. Its scrambling is dominated
by single-body terms in the MBL effective Hamiltonian, which is
different from the MBL cases. Therefore marginal MBL systems have
a different butterfly light cone scaling $\ln t_\text{scr}\sim
d_{WV}^{1/2}$. In this paper, we focused on the case where $W$ and
$V$ are both local operators. Our calculation can be generalized
to generic operators over regions of finite lengths.

\emph{Acknowledgement} --- We would like to thank Yingfei Gu, Xiao
Chen, Xiao-Liang Qi, Yichen Huang, and Yu Chen for inspiring
discussions. The authors are supported by the David and Lucile
Packard Foundation and NSF Grant No. DMR-1151208. We acknowledge
support from the Center for Scientific Computing from the CNSI,
MRL: an NSF MRSEC (DMR-1121053) and NSF CNS-0960316.

\bibliography{OTOC}

\begin{thebibliography}{70}
\expandafter\ifx\csname natexlab\endcsname\relax\def\natexlab#1{#1}\fi
\expandafter\ifx\csname bibnamefont\endcsname\relax
  \def\bibnamefont#1{#1}\fi
\expandafter\ifx\csname bibfnamefont\endcsname\relax
  \def\bibfnamefont#1{#1}\fi
\expandafter\ifx\csname citenamefont\endcsname\relax
  \def\citenamefont#1{#1}\fi
\expandafter\ifx\csname url\endcsname\relax
  \def\url#1{\texttt{#1}}\fi
\expandafter\ifx\csname urlprefix\endcsname\relax\def\urlprefix{URL }\fi
\providecommand{\bibinfo}[2]{#2}
\providecommand{\eprint}[2][]{\url{#2}}

\bibitem[{\citenamefont{Larkin}(1969)}]{Larkin:1969kn}
\bibinfo{author}{\bibfnamefont{Y.~N.} \bibnamefont{Larkin},
  \bibfnamefont{A.;~Ovchinnikov}}, \bibinfo{journal}{Sov. Phys. JETP}
  \textbf{\bibinfo{volume}{28}}, \bibinfo{pages}{1200} (\bibinfo{year}{1969}).

\bibitem[{\citenamefont{{Shenker} and
  {Stanford}}(2014{\natexlab{a}})}]{Shenker:2014fk}
\bibinfo{author}{\bibfnamefont{S.~H.} \bibnamefont{{Shenker}}}
  \bibnamefont{and}
  \bibinfo{author}{\bibfnamefont{D.}~\bibnamefont{{Stanford}}},
  \bibinfo{journal}{Journal of High Energy Physics}
  \textbf{\bibinfo{volume}{3}}, \bibinfo{eid}{67}
  (\bibinfo{year}{2014}{\natexlab{a}}), \eprint{1306.0622}.

\bibitem[{\citenamefont{{Kitaev}}(2014)}]{Kitaev:2014it}
\bibinfo{author}{\bibfnamefont{A.~Y.} \bibnamefont{{Kitaev}}}
  (\bibinfo{year}{2014}), \bibinfo{note}{talk at the Fundamental Physics Prize
  Symposium}.

\bibitem[{\citenamefont{{Shenker} and
  {Stanford}}(2014{\natexlab{b}})}]{Shenker:2014ij}
\bibinfo{author}{\bibfnamefont{S.~H.} \bibnamefont{{Shenker}}}
  \bibnamefont{and}
  \bibinfo{author}{\bibfnamefont{D.}~\bibnamefont{{Stanford}}},
  \bibinfo{journal}{Journal of High Energy Physics}
  \textbf{\bibinfo{volume}{12}}, \bibinfo{eid}{46}
  (\bibinfo{year}{2014}{\natexlab{b}}), \eprint{1312.3296}.

\bibitem[{\citenamefont{{Shenker} and {Stanford}}(2015)}]{Shenker:2015ph}
\bibinfo{author}{\bibfnamefont{S.~H.} \bibnamefont{{Shenker}}}
  \bibnamefont{and}
  \bibinfo{author}{\bibfnamefont{D.}~\bibnamefont{{Stanford}}},
  \bibinfo{journal}{Journal of High Energy Physics}
  \textbf{\bibinfo{volume}{5}}, \bibinfo{eid}{132} (\bibinfo{year}{2015}),
  \eprint{1412.6087}.

\bibitem[{\citenamefont{{Roberts} et~al.}(2015)\citenamefont{{Roberts},
  {Stanford}, and {Susskind}}}]{Roberts:2015oq}
\bibinfo{author}{\bibfnamefont{D.~A.} \bibnamefont{{Roberts}}},
  \bibinfo{author}{\bibfnamefont{D.}~\bibnamefont{{Stanford}}},
  \bibnamefont{and}
  \bibinfo{author}{\bibfnamefont{L.}~\bibnamefont{{Susskind}}},
  \bibinfo{journal}{Journal of High Energy Physics}
  \textbf{\bibinfo{volume}{3}}, \bibinfo{eid}{51} (\bibinfo{year}{2015}),
  \eprint{1409.8180}.

\bibitem[{\citenamefont{{Roberts} and {Stanford}}(2015)}]{Roberts:2015fc}
\bibinfo{author}{\bibfnamefont{D.~A.} \bibnamefont{{Roberts}}}
  \bibnamefont{and}
  \bibinfo{author}{\bibfnamefont{D.}~\bibnamefont{{Stanford}}},
  \bibinfo{journal}{Physical Review Letters} \textbf{\bibinfo{volume}{115}},
  \bibinfo{eid}{131603} (\bibinfo{year}{2015}), \eprint{1412.5123}.

\bibitem[{\citenamefont{{Kitaev}}(2015)}]{Kitaev:2015uj}
\bibinfo{author}{\bibfnamefont{A.}~\bibnamefont{{Kitaev}}}
  (\bibinfo{year}{2015}), \bibinfo{note}{talk at KITP Program: Entanglement in
  Strongly-Correlated Quantum Matter},
  \urlprefix\url{http://online.kitp.ucsb.edu/online/entangled15/kitaev/}.

\bibitem[{\citenamefont{{Maldacena} and {Stanford}}(2016)}]{Maldacena:2016hh}
\bibinfo{author}{\bibfnamefont{J.}~\bibnamefont{{Maldacena}}} \bibnamefont{and}
  \bibinfo{author}{\bibfnamefont{D.}~\bibnamefont{{Stanford}}},
  \bibinfo{journal}{ArXiv e-prints}  (\bibinfo{year}{2016}),
  \eprint{1604.07818}.

\bibitem[{\citenamefont{{Maldacena}
  et~al.}(2016{\natexlab{a}})\citenamefont{{Maldacena}, {Stanford}, and
  {Yang}}}]{Maldacena:2016bx}
\bibinfo{author}{\bibfnamefont{J.}~\bibnamefont{{Maldacena}}},
  \bibinfo{author}{\bibfnamefont{D.}~\bibnamefont{{Stanford}}},
  \bibnamefont{and} \bibinfo{author}{\bibfnamefont{Z.}~\bibnamefont{{Yang}}},
  \bibinfo{journal}{ArXiv e-prints}  (\bibinfo{year}{2016}{\natexlab{a}}),
  \eprint{1606.01857}.

\bibitem[{\citenamefont{{Maldacena}
  et~al.}(2016{\natexlab{b}})\citenamefont{{Maldacena}, {Shenker}, and
  {Stanford}}}]{Maldacena:2016xy}
\bibinfo{author}{\bibfnamefont{J.}~\bibnamefont{{Maldacena}}},
  \bibinfo{author}{\bibfnamefont{S.~H.} \bibnamefont{{Shenker}}},
  \bibnamefont{and}
  \bibinfo{author}{\bibfnamefont{D.}~\bibnamefont{{Stanford}}},
  \bibinfo{journal}{Journal of High Energy Physics}
  \textbf{\bibinfo{volume}{8}}, \bibinfo{pages}{106}
  (\bibinfo{year}{2016}{\natexlab{b}}), \eprint{1503.01409}.

\bibitem[{\citenamefont{{Hosur} et~al.}(2016)\citenamefont{{Hosur}, {Qi},
  {Roberts}, and {Yoshida}}}]{Hosur:2016fy}
\bibinfo{author}{\bibfnamefont{P.}~\bibnamefont{{Hosur}}},
  \bibinfo{author}{\bibfnamefont{X.-L.} \bibnamefont{{Qi}}},
  \bibinfo{author}{\bibfnamefont{D.~A.} \bibnamefont{{Roberts}}},
  \bibnamefont{and}
  \bibinfo{author}{\bibfnamefont{B.}~\bibnamefont{{Yoshida}}},
  \bibinfo{journal}{Journal of High Energy Physics}
  \textbf{\bibinfo{volume}{2}}, \bibinfo{eid}{4} (\bibinfo{year}{2016}),
  \eprint{1511.04021}.

\bibitem[{\citenamefont{{Gu} and {Qi}}(2016)}]{Gu:2016hi}
\bibinfo{author}{\bibfnamefont{Y.}~\bibnamefont{{Gu}}} \bibnamefont{and}
  \bibinfo{author}{\bibfnamefont{X.-L.} \bibnamefont{{Qi}}},
  \bibinfo{journal}{Journal of High Energy Physics}
  \textbf{\bibinfo{volume}{8}}, \bibinfo{eid}{129} (\bibinfo{year}{2016}),
  \eprint{1602.06543}.

\bibitem[{\citenamefont{{Fu} and {Sachdev}}(2016)}]{Fu:2016hs}
\bibinfo{author}{\bibfnamefont{W.}~\bibnamefont{{Fu}}} \bibnamefont{and}
  \bibinfo{author}{\bibfnamefont{S.}~\bibnamefont{{Sachdev}}},
  \bibinfo{journal}{\prb} \textbf{\bibinfo{volume}{94}}, \bibinfo{eid}{035135}
  (\bibinfo{year}{2016}), \eprint{1603.05246}.

\bibitem[{\citenamefont{{Swingle} et~al.}(2016)\citenamefont{{Swingle},
  {Bentsen}, {Schleier-Smith}, and {Hayden}}}]{Swingle:2016oe}
\bibinfo{author}{\bibfnamefont{B.}~\bibnamefont{{Swingle}}},
  \bibinfo{author}{\bibfnamefont{G.}~\bibnamefont{{Bentsen}}},
  \bibinfo{author}{\bibfnamefont{M.}~\bibnamefont{{Schleier-Smith}}},
  \bibnamefont{and} \bibinfo{author}{\bibfnamefont{P.}~\bibnamefont{{Hayden}}},
  \bibinfo{journal}{ArXiv e-prints}  (\bibinfo{year}{2016}),
  \eprint{1602.06271}.

\bibitem[{\citenamefont{{Zhu} et~al.}(2016)\citenamefont{{Zhu}, {Hafezi}, and
  {Grover}}}]{Zhu:2016cr}
\bibinfo{author}{\bibfnamefont{G.}~\bibnamefont{{Zhu}}},
  \bibinfo{author}{\bibfnamefont{M.}~\bibnamefont{{Hafezi}}}, \bibnamefont{and}
  \bibinfo{author}{\bibfnamefont{T.}~\bibnamefont{{Grover}}},
  \bibinfo{journal}{ArXiv e-prints}  (\bibinfo{year}{2016}),
  \eprint{1607.00079}.

\bibitem[{\citenamefont{{Yao} et~al.}(2016)\citenamefont{{Yao}, {Grusdt},
  {Swingle}, {Lukin}, {Stamper-Kurn}, {Moore}, and {Demler}}}]{Yao:2016xp}
\bibinfo{author}{\bibfnamefont{N.~Y.} \bibnamefont{{Yao}}},
  \bibinfo{author}{\bibfnamefont{F.}~\bibnamefont{{Grusdt}}},
  \bibinfo{author}{\bibfnamefont{B.}~\bibnamefont{{Swingle}}},
  \bibinfo{author}{\bibfnamefont{M.~D.} \bibnamefont{{Lukin}}},
  \bibinfo{author}{\bibfnamefont{D.~M.} \bibnamefont{{Stamper-Kurn}}},
  \bibinfo{author}{\bibfnamefont{J.~E.} \bibnamefont{{Moore}}},
  \bibnamefont{and} \bibinfo{author}{\bibfnamefont{E.~A.}
  \bibnamefont{{Demler}}}, \bibinfo{journal}{ArXiv e-prints}
  (\bibinfo{year}{2016}), \eprint{1607.01801}.

\bibitem[{\citenamefont{Page}(1993)}]{Page:1993fv}
\bibinfo{author}{\bibfnamefont{D.~N.} \bibnamefont{Page}},
  \bibinfo{journal}{Phys. Rev. Lett.} \textbf{\bibinfo{volume}{71}},
  \bibinfo{pages}{1291} (\bibinfo{year}{1993}).

\bibitem[{\citenamefont{{Hayden} and {Preskill}}(2007)}]{Hayden:2007uf}
\bibinfo{author}{\bibfnamefont{P.}~\bibnamefont{{Hayden}}} \bibnamefont{and}
  \bibinfo{author}{\bibfnamefont{J.}~\bibnamefont{{Preskill}}},
  \bibinfo{journal}{Journal of High Energy Physics}
  \textbf{\bibinfo{volume}{9}}, \bibinfo{eid}{120} (\bibinfo{year}{2007}),
  \eprint{0708.4025}.

\bibitem[{\citenamefont{{Sekino} and {Susskind}}(2008)}]{Sekino:2008pv}
\bibinfo{author}{\bibfnamefont{Y.}~\bibnamefont{{Sekino}}} \bibnamefont{and}
  \bibinfo{author}{\bibfnamefont{L.}~\bibnamefont{{Susskind}}},
  \bibinfo{journal}{Journal of High Energy Physics}
  \textbf{\bibinfo{volume}{10}}, \bibinfo{eid}{065} (\bibinfo{year}{2008}),
  \eprint{0808.2096}.

\bibitem[{\citenamefont{{Lashkari} et~al.}(2013)\citenamefont{{Lashkari},
  {Stanford}, {Hastings}, {Osborne}, and {Hayden}}}]{Lashkari:2013mj}
\bibinfo{author}{\bibfnamefont{N.}~\bibnamefont{{Lashkari}}},
  \bibinfo{author}{\bibfnamefont{D.}~\bibnamefont{{Stanford}}},
  \bibinfo{author}{\bibfnamefont{M.}~\bibnamefont{{Hastings}}},
  \bibinfo{author}{\bibfnamefont{T.}~\bibnamefont{{Osborne}}},
  \bibnamefont{and} \bibinfo{author}{\bibfnamefont{P.}~\bibnamefont{{Hayden}}},
  \bibinfo{journal}{Journal of High Energy Physics}
  \textbf{\bibinfo{volume}{4}}, \bibinfo{eid}{22} (\bibinfo{year}{2013}),
  \eprint{1111.6580}.

\bibitem[{\citenamefont{{Caputa} et~al.}(2016)\citenamefont{{Caputa},
  {Numasawa}, and {Veliz-Osorio}}}]{Caputa:2016ct}
\bibinfo{author}{\bibfnamefont{P.}~\bibnamefont{{Caputa}}},
  \bibinfo{author}{\bibfnamefont{T.}~\bibnamefont{{Numasawa}}},
  \bibnamefont{and}
  \bibinfo{author}{\bibfnamefont{A.}~\bibnamefont{{Veliz-Osorio}}},
  \bibinfo{journal}{ArXiv e-prints}  (\bibinfo{year}{2016}),
  \eprint{1602.06542}.

\bibitem[{\citenamefont{{Huang} et~al.}(2016)\citenamefont{{Huang}, {Zhang},
  and {Chen}}}]{Huang:2016wk}
\bibinfo{author}{\bibfnamefont{Y.}~\bibnamefont{{Huang}}},
  \bibinfo{author}{\bibfnamefont{Y.-L.} \bibnamefont{{Zhang}}},
  \bibnamefont{and} \bibinfo{author}{\bibfnamefont{X.}~\bibnamefont{{Chen}}},
  \bibinfo{journal}{ArXiv e-prints}  (\bibinfo{year}{2016}),
  \eprint{1608.01091}.

\bibitem[{\citenamefont{{Fan} et~al.}(2016)\citenamefont{{Fan}, {Zhang},
  {Shen}, and {Zhai}}}]{Fan:2016oy}
\bibinfo{author}{\bibfnamefont{R.}~\bibnamefont{{Fan}}},
  \bibinfo{author}{\bibfnamefont{P.}~\bibnamefont{{Zhang}}},
  \bibinfo{author}{\bibfnamefont{H.}~\bibnamefont{{Shen}}}, \bibnamefont{and}
  \bibinfo{author}{\bibfnamefont{H.}~\bibnamefont{{Zhai}}},
  \bibinfo{journal}{ArXiv e-prints}  (\bibinfo{year}{2016}),
  \eprint{1608.01914}.

\bibitem[{\citenamefont{{Swingle} and {Chowdhury}}(2016)}]{Swingle:2016rw}
\bibinfo{author}{\bibfnamefont{B.}~\bibnamefont{{Swingle}}} \bibnamefont{and}
  \bibinfo{author}{\bibfnamefont{D.}~\bibnamefont{{Chowdhury}}},
  \bibinfo{journal}{ArXiv e-prints}  (\bibinfo{year}{2016}),
  \eprint{1608.03280}.

\bibitem[{\citenamefont{{Chen}}(2016)}]{Chen:2016ec}
\bibinfo{author}{\bibfnamefont{Y.}~\bibnamefont{{Chen}}},
  \bibinfo{journal}{ArXiv e-prints}  (\bibinfo{year}{2016}),
  \eprint{1608.02765}.

\bibitem[{\citenamefont{{He} and {Lu}}(2016)}]{He:2016bf}
\bibinfo{author}{\bibfnamefont{R.-Q.} \bibnamefont{{He}}} \bibnamefont{and}
  \bibinfo{author}{\bibfnamefont{Z.-Y.} \bibnamefont{{Lu}}},
  \bibinfo{journal}{ArXiv e-prints}  (\bibinfo{year}{2016}),
  \eprint{1608.03586}.

\bibitem[{\citenamefont{{Chen} et~al.}(2016)\citenamefont{{Chen}, {Zhou},
  {Huse}, and {Fradkin}}}]{Chen:2016bx}
\bibinfo{author}{\bibfnamefont{X.}~\bibnamefont{{Chen}}},
  \bibinfo{author}{\bibfnamefont{T.}~\bibnamefont{{Zhou}}},
  \bibinfo{author}{\bibfnamefont{D.~A.} \bibnamefont{{Huse}}},
  \bibnamefont{and}
  \bibinfo{author}{\bibfnamefont{E.}~\bibnamefont{{Fradkin}}},
  \bibinfo{journal}{ArXiv e-prints}  (\bibinfo{year}{2016}),
  \eprint{1610.00220}.

\bibitem[{\citenamefont{Berkovits and Avishai}(1996)}]{Avishai:1996qe}
\bibinfo{author}{\bibfnamefont{R.}~\bibnamefont{Berkovits}} \bibnamefont{and}
  \bibinfo{author}{\bibfnamefont{Y.}~\bibnamefont{Avishai}},
  \bibinfo{journal}{Journal of Physics: Condensed Matter}
  \textbf{\bibinfo{volume}{8}}, \bibinfo{pages}{389} (\bibinfo{year}{1996}).

\bibitem[{\citenamefont{Berkovits and Avishai}(1997)}]{Avishai:1997cs}
\bibinfo{author}{\bibfnamefont{R.}~\bibnamefont{Berkovits}} \bibnamefont{and}
  \bibinfo{author}{\bibfnamefont{Y.}~\bibnamefont{Avishai}},
  \bibinfo{journal}{arXiv preprint cond-mat/9707066}  (\bibinfo{year}{1997}).

\bibitem[{\citenamefont{{Gornyi} et~al.}(2005)\citenamefont{{Gornyi}, {Mirlin},
  and {Polyakov}}}]{GMP:2005}
\bibinfo{author}{\bibfnamefont{I.~V.} \bibnamefont{{Gornyi}}},
  \bibinfo{author}{\bibfnamefont{A.~D.} \bibnamefont{{Mirlin}}},
  \bibnamefont{and} \bibinfo{author}{\bibfnamefont{D.~G.}
  \bibnamefont{{Polyakov}}}, \bibinfo{journal}{Physical Review Letters}
  \textbf{\bibinfo{volume}{95}}, \bibinfo{eid}{206603} (\bibinfo{year}{2005}),
  \eprint{cond-mat/0506411}.

\bibitem[{\citenamefont{{Basko} et~al.}(2006)\citenamefont{{Basko}, {Aleiner},
  and {Altshuler}}}]{BAA:2006}
\bibinfo{author}{\bibfnamefont{D.~M.} \bibnamefont{{Basko}}},
  \bibinfo{author}{\bibfnamefont{I.~L.} \bibnamefont{{Aleiner}}},
  \bibnamefont{and} \bibinfo{author}{\bibfnamefont{B.~L.}
  \bibnamefont{{Altshuler}}}, \bibinfo{journal}{Annals of Physics}
  \textbf{\bibinfo{volume}{321}}, \bibinfo{pages}{1126} (\bibinfo{year}{2006}),
  \eprint{cond-mat/0506617}.

\bibitem[{\citenamefont{{Imbrie}}(2014)}]{Imbrie:2014}
\bibinfo{author}{\bibfnamefont{J.~Z.} \bibnamefont{{Imbrie}}},
  \bibinfo{journal}{ArXiv e-prints}  (\bibinfo{year}{2014}),
  \eprint{1403.7837}.

\bibitem[{\citenamefont{{Nandkishore} and {Huse}}(2015)}]{Huse:2015rv}
\bibinfo{author}{\bibfnamefont{R.}~\bibnamefont{{Nandkishore}}}
  \bibnamefont{and} \bibinfo{author}{\bibfnamefont{D.~A.}
  \bibnamefont{{Huse}}}, \bibinfo{journal}{Annual Review of Condensed Matter
  Physics} \textbf{\bibinfo{volume}{6}}, \bibinfo{pages}{15}
  (\bibinfo{year}{2015}), \eprint{1404.0686}.

\bibitem[{\citenamefont{{Bardarson} et~al.}(2012)\citenamefont{{Bardarson},
  {Pollmann}, and {Moore}}}]{Moore:2012ge}
\bibinfo{author}{\bibfnamefont{J.~H.} \bibnamefont{{Bardarson}}},
  \bibinfo{author}{\bibfnamefont{F.}~\bibnamefont{{Pollmann}}},
  \bibnamefont{and} \bibinfo{author}{\bibfnamefont{J.~E.}
  \bibnamefont{{Moore}}}, \bibinfo{journal}{Physical Review Letters}
  \textbf{\bibinfo{volume}{109}}, \bibinfo{eid}{017202} (\bibinfo{year}{2012}),
  \eprint{1202.5532}.

\bibitem[{\citenamefont{{DeChiara} et~al.}(2006)\citenamefont{{DeChiara},
  {Montangero}, {Calabrese}, and {Fazio}}}]{DeChiara:2006zf}
\bibinfo{author}{\bibfnamefont{G.}~\bibnamefont{{DeChiara}}},
  \bibinfo{author}{\bibfnamefont{S.}~\bibnamefont{{Montangero}}},
  \bibinfo{author}{\bibfnamefont{P.}~\bibnamefont{{Calabrese}}},
  \bibnamefont{and} \bibinfo{author}{\bibfnamefont{R.}~\bibnamefont{{Fazio}}},
  \bibinfo{journal}{Journal of Statistical Mechanics: Theory and Experiment}
  \textbf{\bibinfo{volume}{3}}, \bibinfo{pages}{03001} (\bibinfo{year}{2006}),
  \eprint{cond-mat/0512586}.

\bibitem[{\citenamefont{{{\v Z}nidari{\v c}} et~al.}(2008)\citenamefont{{{\v
  Z}nidari{\v c}}, {Prosen}, and {Prelov{\v s}ek}}}]{Znidaric:2008he}
\bibinfo{author}{\bibfnamefont{M.}~\bibnamefont{{{\v Z}nidari{\v c}}}},
  \bibinfo{author}{\bibfnamefont{T.}~\bibnamefont{{Prosen}}}, \bibnamefont{and}
  \bibinfo{author}{\bibfnamefont{P.}~\bibnamefont{{Prelov{\v s}ek}}},
  \bibinfo{journal}{\prb} \textbf{\bibinfo{volume}{77}}, \bibinfo{eid}{064426}
  (\bibinfo{year}{2008}), \eprint{0706.2539}.

\bibitem[{\citenamefont{{Serbyn}
  et~al.}(2013{\natexlab{a}})\citenamefont{{Serbyn}, {Papi{\'c}}, and
  {Abanin}}}]{Abanin:2013ta}
\bibinfo{author}{\bibfnamefont{M.}~\bibnamefont{{Serbyn}}},
  \bibinfo{author}{\bibfnamefont{Z.}~\bibnamefont{{Papi{\'c}}}},
  \bibnamefont{and} \bibinfo{author}{\bibfnamefont{D.~A.}
  \bibnamefont{{Abanin}}}, \bibinfo{journal}{Physical Review Letters}
  \textbf{\bibinfo{volume}{110}}, \bibinfo{eid}{260601}
  (\bibinfo{year}{2013}{\natexlab{a}}), \eprint{1304.4605}.

\bibitem[{\citenamefont{{Nanduri} et~al.}(2014)\citenamefont{{Nanduri}, {Kim},
  and {Huse}}}]{Nanduri:2014jh}
\bibinfo{author}{\bibfnamefont{A.}~\bibnamefont{{Nanduri}}},
  \bibinfo{author}{\bibfnamefont{H.}~\bibnamefont{{Kim}}}, \bibnamefont{and}
  \bibinfo{author}{\bibfnamefont{D.~A.} \bibnamefont{{Huse}}},
  \bibinfo{journal}{\prb} \textbf{\bibinfo{volume}{90}}, \bibinfo{eid}{064201}
  (\bibinfo{year}{2014}), \eprint{1404.5216}.

\bibitem[{\citenamefont{{Shen} et~al.}(2016)\citenamefont{{Shen}, {Zhang},
  {Fan}, and {Zhai}}}]{Shen:2016ov}
\bibinfo{author}{\bibfnamefont{H.}~\bibnamefont{{Shen}}},
  \bibinfo{author}{\bibfnamefont{P.}~\bibnamefont{{Zhang}}},
  \bibinfo{author}{\bibfnamefont{R.}~\bibnamefont{{Fan}}}, \bibnamefont{and}
  \bibinfo{author}{\bibfnamefont{H.}~\bibnamefont{{Zhai}}},
  \bibinfo{journal}{ArXiv e-prints}  (\bibinfo{year}{2016}),
  \eprint{1608.02438}.

\bibitem[{\citenamefont{{Nandkishore} and {Potter}}(2014)}]{Potter:2014mh}
\bibinfo{author}{\bibfnamefont{R.}~\bibnamefont{{Nandkishore}}}
  \bibnamefont{and} \bibinfo{author}{\bibfnamefont{A.~C.}
  \bibnamefont{{Potter}}}, \bibinfo{journal}{\prb}
  \textbf{\bibinfo{volume}{90}}, \bibinfo{eid}{195115} (\bibinfo{year}{2014}),
  \eprint{1406.0847}.

\bibitem[{\citenamefont{{You} et~al.}(2015)\citenamefont{{You}, {Qi}, and
  {Xu}}}]{You:2015sb}
\bibinfo{author}{\bibfnamefont{Y.-Z.} \bibnamefont{{You}}},
  \bibinfo{author}{\bibfnamefont{X.-L.} \bibnamefont{{Qi}}}, \bibnamefont{and}
  \bibinfo{author}{\bibfnamefont{C.}~\bibnamefont{{Xu}}},
  \bibinfo{journal}{ArXiv e-prints}  (\bibinfo{year}{2015}),
  \eprint{1508.03635}.

\bibitem[{\citenamefont{{Slagle} et~al.}(2016)\citenamefont{{Slagle}, {You},
  and {Xu}}}]{Slagle:2016mk}
\bibinfo{author}{\bibfnamefont{K.}~\bibnamefont{{Slagle}}},
  \bibinfo{author}{\bibfnamefont{Y.-Z.} \bibnamefont{{You}}}, \bibnamefont{and}
  \bibinfo{author}{\bibfnamefont{C.}~\bibnamefont{{Xu}}},
  \bibinfo{journal}{\prb} \textbf{\bibinfo{volume}{94}}, \bibinfo{eid}{014205}
  (\bibinfo{year}{2016}), \eprint{1604.04283}.

\bibitem[{\citenamefont{{Vosk} and {Altman}}(2013)}]{Altman:2013rg}
\bibinfo{author}{\bibfnamefont{R.}~\bibnamefont{{Vosk}}} \bibnamefont{and}
  \bibinfo{author}{\bibfnamefont{E.}~\bibnamefont{{Altman}}},
  \bibinfo{journal}{Physical Review Letters} \textbf{\bibinfo{volume}{110}},
  \bibinfo{eid}{067204} (\bibinfo{year}{2013}), \eprint{1205.0026}.

\bibitem[{\citenamefont{{Swingle}}(2013)}]{Swingle:2013oy}
\bibinfo{author}{\bibfnamefont{B.}~\bibnamefont{{Swingle}}},
  \bibinfo{journal}{ArXiv e-prints}  (\bibinfo{year}{2013}),
  \eprint{1307.0507}.

\bibitem[{\citenamefont{{Refael} and {Altman}}(2013)}]{Refael:2013du}
\bibinfo{author}{\bibfnamefont{G.}~\bibnamefont{{Refael}}} \bibnamefont{and}
  \bibinfo{author}{\bibfnamefont{E.}~\bibnamefont{{Altman}}},
  \bibinfo{journal}{Comptes Rendus Physique} \textbf{\bibinfo{volume}{14}},
  \bibinfo{pages}{725} (\bibinfo{year}{2013}), \eprint{1402.6008}.

\bibitem[{\citenamefont{{Pekker} et~al.}(2014)\citenamefont{{Pekker}, {Refael},
  {Altman}, {Demler}, and {Oganesyan}}}]{Altman:2014hg}
\bibinfo{author}{\bibfnamefont{D.}~\bibnamefont{{Pekker}}},
  \bibinfo{author}{\bibfnamefont{G.}~\bibnamefont{{Refael}}},
  \bibinfo{author}{\bibfnamefont{E.}~\bibnamefont{{Altman}}},
  \bibinfo{author}{\bibfnamefont{E.}~\bibnamefont{{Demler}}}, \bibnamefont{and}
  \bibinfo{author}{\bibfnamefont{V.}~\bibnamefont{{Oganesyan}}},
  \bibinfo{journal}{Physical Review X} \textbf{\bibinfo{volume}{4}},
  \bibinfo{eid}{011052} (\bibinfo{year}{2014}), \eprint{1307.3253}.

\bibitem[{\citenamefont{{Vasseur}
  et~al.}(2015{\natexlab{a}})\citenamefont{{Vasseur}, {Potter}, and
  {Parameswaran}}}]{Potter:2015th}
\bibinfo{author}{\bibfnamefont{R.}~\bibnamefont{{Vasseur}}},
  \bibinfo{author}{\bibfnamefont{A.~C.} \bibnamefont{{Potter}}},
  \bibnamefont{and} \bibinfo{author}{\bibfnamefont{S.~A.}
  \bibnamefont{{Parameswaran}}}, \bibinfo{journal}{Physical Review Letters}
  \textbf{\bibinfo{volume}{114}}, \bibinfo{eid}{217201}
  (\bibinfo{year}{2015}{\natexlab{a}}), \eprint{1410.6165}.

\bibitem[{\citenamefont{{Vasseur}
  et~al.}(2015{\natexlab{b}})\citenamefont{{Vasseur}, {Friedman},
  {Parameswaran}, and {Potter}}}]{Vasseur:2015ys}
\bibinfo{author}{\bibfnamefont{R.}~\bibnamefont{{Vasseur}}},
  \bibinfo{author}{\bibfnamefont{A.~J.} \bibnamefont{{Friedman}}},
  \bibinfo{author}{\bibfnamefont{S.~A.} \bibnamefont{{Parameswaran}}},
  \bibnamefont{and} \bibinfo{author}{\bibfnamefont{A.~C.}
  \bibnamefont{{Potter}}}, \bibinfo{journal}{ArXiv e-prints}
  (\bibinfo{year}{2015}{\natexlab{b}}), \eprint{1510.04282}.

\bibitem[{\citenamefont{{Serbyn}
  et~al.}(2013{\natexlab{b}})\citenamefont{{Serbyn}, {Papi{\'c}}, and
  {Abanin}}}]{Abanin:2013lc}
\bibinfo{author}{\bibfnamefont{M.}~\bibnamefont{{Serbyn}}},
  \bibinfo{author}{\bibfnamefont{Z.}~\bibnamefont{{Papi{\'c}}}},
  \bibnamefont{and} \bibinfo{author}{\bibfnamefont{D.~A.}
  \bibnamefont{{Abanin}}}, \bibinfo{journal}{Physical Review Letters}
  \textbf{\bibinfo{volume}{111}}, \bibinfo{eid}{127201}
  (\bibinfo{year}{2013}{\natexlab{b}}), \eprint{1305.5554}.

\bibitem[{\citenamefont{{Huse} et~al.}(2014)\citenamefont{{Huse},
  {Nandkishore}, and {Oganesyan}}}]{Huse:2014ec}
\bibinfo{author}{\bibfnamefont{D.~A.} \bibnamefont{{Huse}}},
  \bibinfo{author}{\bibfnamefont{R.}~\bibnamefont{{Nandkishore}}},
  \bibnamefont{and}
  \bibinfo{author}{\bibfnamefont{V.}~\bibnamefont{{Oganesyan}}},
  \bibinfo{journal}{\prb} \textbf{\bibinfo{volume}{90}}, \bibinfo{eid}{174202}
  (\bibinfo{year}{2014}), \eprint{1305.4915}.

\bibitem[{\citenamefont{{Kim} et~al.}(2014)\citenamefont{{Kim}, {Chandran}, and
  {Abanin}}}]{Kim:2014zj}
\bibinfo{author}{\bibfnamefont{I.~H.} \bibnamefont{{Kim}}},
  \bibinfo{author}{\bibfnamefont{A.}~\bibnamefont{{Chandran}}},
  \bibnamefont{and} \bibinfo{author}{\bibfnamefont{D.~A.}
  \bibnamefont{{Abanin}}}, \bibinfo{journal}{ArXiv e-prints}
  (\bibinfo{year}{2014}), \eprint{1412.3073}.

\bibitem[{\citenamefont{{Chandran} et~al.}(2015)\citenamefont{{Chandran},
  {Kim}, {Vidal}, and {Abanin}}}]{Abanin:2015io}
\bibinfo{author}{\bibfnamefont{A.}~\bibnamefont{{Chandran}}},
  \bibinfo{author}{\bibfnamefont{I.~H.} \bibnamefont{{Kim}}},
  \bibinfo{author}{\bibfnamefont{G.}~\bibnamefont{{Vidal}}}, \bibnamefont{and}
  \bibinfo{author}{\bibfnamefont{D.~A.} \bibnamefont{{Abanin}}},
  \bibinfo{journal}{\prb} \textbf{\bibinfo{volume}{91}}, \bibinfo{eid}{085425}
  (\bibinfo{year}{2015}), \eprint{1407.8480}.

\bibitem[{\citenamefont{{Rademaker}}(2015)}]{Rademaker:2015ve}
\bibinfo{author}{\bibfnamefont{L.}~\bibnamefont{{Rademaker}}},
  \bibinfo{journal}{ArXiv e-prints}  (\bibinfo{year}{2015}),
  \eprint{1507.07276}.

\bibitem[{\citenamefont{Srednicki}(1994)}]{Srednicki:1994ns}
\bibinfo{author}{\bibfnamefont{M.}~\bibnamefont{Srednicki}},
  \bibinfo{journal}{Phys. Rev. E} \textbf{\bibinfo{volume}{50}},
  \bibinfo{pages}{888} (\bibinfo{year}{1994}).

\bibitem[{\citenamefont{Santos and Rigol}(2010)}]{Rigol:2010ls}
\bibinfo{author}{\bibfnamefont{L.~F.} \bibnamefont{Santos}} \bibnamefont{and}
  \bibinfo{author}{\bibfnamefont{M.}~\bibnamefont{Rigol}},
  \bibinfo{journal}{Phys. Rev. E} \textbf{\bibinfo{volume}{81}},
  \bibinfo{pages}{036206} (\bibinfo{year}{2010}).

\bibitem[{\citenamefont{{Hartnoll}}(2015)}]{Hartnoll:2015ef}
\bibinfo{author}{\bibfnamefont{S.~A.} \bibnamefont{{Hartnoll}}},
  \bibinfo{journal}{Nature Physics} \textbf{\bibinfo{volume}{11}},
  \bibinfo{pages}{54} (\bibinfo{year}{2015}), \eprint{1405.3651}.

\bibitem[{\citenamefont{{Blake}}(2016{\natexlab{a}})}]{Blake:2016ee}
\bibinfo{author}{\bibfnamefont{M.}~\bibnamefont{{Blake}}},
  \bibinfo{journal}{ArXiv e-prints}  (\bibinfo{year}{2016}{\natexlab{a}}),
  \eprint{1603.08510}.

\bibitem[{\citenamefont{{Blake}}(2016{\natexlab{b}})}]{Blake:2016xw}
\bibinfo{author}{\bibfnamefont{M.}~\bibnamefont{{Blake}}},
  \bibinfo{journal}{ArXiv e-prints}  (\bibinfo{year}{2016}{\natexlab{b}}),
  \eprint{1604.01754}.

\bibitem[{\citenamefont{{Gu} et~al.}(2016)\citenamefont{{Gu}, {Qi}, and
  {Stanford}}}]{Gu:2016zm}
\bibinfo{author}{\bibfnamefont{Y.}~\bibnamefont{{Gu}}},
  \bibinfo{author}{\bibfnamefont{X.-L.} \bibnamefont{{Qi}}}, \bibnamefont{and}
  \bibinfo{author}{\bibfnamefont{D.}~\bibnamefont{{Stanford}}},
  \bibinfo{journal}{ArXiv e-prints}  (\bibinfo{year}{2016}),
  \eprint{1609.07832}.

\bibitem[{\citenamefont{Chen et~al.}(2003)\citenamefont{Chen, Härdle, and
  Li}}]{ChenErrors}
\bibinfo{author}{\bibfnamefont{S.~X.} \bibnamefont{Chen}},
  \bibinfo{author}{\bibfnamefont{W.}~\bibnamefont{Härdle}}, \bibnamefont{and}
  \bibinfo{author}{\bibfnamefont{M.}~\bibnamefont{Li}},
  \bibinfo{journal}{Journal of the Royal Statistical Society: Series B
  (Statistical Methodology)} \textbf{\bibinfo{volume}{65}},
  \bibinfo{pages}{663} (\bibinfo{year}{2003}), ISSN \bibinfo{issn}{1467-9868},
  \urlprefix\url{http://dx.doi.org/10.1111/1467-9868.00408}.

\bibitem[{\citenamefont{Efron}(1979)}]{EfronBootstrap}
\bibinfo{author}{\bibfnamefont{B.}~\bibnamefont{Efron}}, \bibinfo{journal}{Ann.
  Statist.} \textbf{\bibinfo{volume}{7}}, \bibinfo{pages}{1}
  (\bibinfo{year}{1979}),
  \urlprefix\url{http://dx.doi.org/10.1214/aos/1176344552}.

\bibitem[{\citenamefont{Fisher}(1992)}]{Fisher:1992is}
\bibinfo{author}{\bibfnamefont{D.~S.} \bibnamefont{Fisher}},
  \bibinfo{journal}{Phys. Rev. Lett.} \textbf{\bibinfo{volume}{69}},
  \bibinfo{pages}{534} (\bibinfo{year}{1992}).

\bibitem[{\citenamefont{Fisher}(1994)}]{Fisher:1994he}
\bibinfo{author}{\bibfnamefont{D.~S.} \bibnamefont{Fisher}},
  \bibinfo{journal}{Phys. Rev. B} \textbf{\bibinfo{volume}{50}},
  \bibinfo{pages}{3799} (\bibinfo{year}{1994}).

\bibitem[{\citenamefont{Fisher}(1995)}]{Fisher:1995cr}
\bibinfo{author}{\bibfnamefont{D.~S.} \bibnamefont{Fisher}},
  \bibinfo{journal}{Phys. Rev. B} \textbf{\bibinfo{volume}{51}},
  \bibinfo{pages}{6411} (\bibinfo{year}{1995}).

\bibitem[{\citenamefont{{Motrunich} et~al.}(2001)\citenamefont{{Motrunich},
  {Damle}, and {Huse}}}]{Huse:2001ez}
\bibinfo{author}{\bibfnamefont{O.}~\bibnamefont{{Motrunich}}},
  \bibinfo{author}{\bibfnamefont{K.}~\bibnamefont{{Damle}}}, \bibnamefont{and}
  \bibinfo{author}{\bibfnamefont{D.~A.} \bibnamefont{{Huse}}},
  \bibinfo{journal}{\prb} \textbf{\bibinfo{volume}{63}}, \bibinfo{eid}{134424}
  (\bibinfo{year}{2001}), \eprint{cond-mat/0005543}.

\bibitem[{\citenamefont{Sinai}(1982)}]{Sinai:1982sf}
\bibinfo{author}{\bibfnamefont{Y.~G.} \bibnamefont{Sinai}},
  \bibinfo{journal}{Theory Probab. Appl.} \textbf{\bibinfo{volume}{27}}
  (\bibinfo{year}{1982}).

\bibitem[{\citenamefont{{Bagrets} et~al.}(2016)\citenamefont{{Bagrets},
  {Altland}, and {Kamenev}}}]{Bagrets:2016kx}
\bibinfo{author}{\bibfnamefont{D.}~\bibnamefont{{Bagrets}}},
  \bibinfo{author}{\bibfnamefont{A.}~\bibnamefont{{Altland}}},
  \bibnamefont{and}
  \bibinfo{author}{\bibfnamefont{A.}~\bibnamefont{{Kamenev}}},
  \bibinfo{journal}{ArXiv e-prints}  (\bibinfo{year}{2016}),
  \eprint{1605.01657}.

\bibitem[{\citenamefont{{Vosk} and {Altman}}(2014)}]{Altman:2014dq}
\bibinfo{author}{\bibfnamefont{R.}~\bibnamefont{{Vosk}}} \bibnamefont{and}
  \bibinfo{author}{\bibfnamefont{E.}~\bibnamefont{{Altman}}},
  \bibinfo{journal}{Physical Review Letters} \textbf{\bibinfo{volume}{112}},
  \bibinfo{eid}{217204} (\bibinfo{year}{2014}), \eprint{1307.3256}.

\bibitem[{\citenamefont{{Gopalakrishnan}
  et~al.}(2015)\citenamefont{{Gopalakrishnan}, {Mueller}, {Khemani}, {Knap},
  {Demler}, and {Huse}}}]{Huse:2015cd}
\bibinfo{author}{\bibfnamefont{S.}~\bibnamefont{{Gopalakrishnan}}},
  \bibinfo{author}{\bibfnamefont{M.}~\bibnamefont{{Mueller}}},
  \bibinfo{author}{\bibfnamefont{V.}~\bibnamefont{{Khemani}}},
  \bibinfo{author}{\bibfnamefont{M.}~\bibnamefont{{Knap}}},
  \bibinfo{author}{\bibfnamefont{E.}~\bibnamefont{{Demler}}}, \bibnamefont{and}
  \bibinfo{author}{\bibfnamefont{D.~A.} \bibnamefont{{Huse}}},
  \bibinfo{journal}{ArXiv e-prints}  (\bibinfo{year}{2015}),
  \eprint{1502.07712}.

\end{thebibliography}
\bibliographystyle{apsrev}

\appendix

\end{document}